\documentclass[12pt]{article}
\usepackage{amsfonts}
\usepackage{amsmath}
\usepackage{amssymb}
\usepackage{graphicx}%
\begin{document}

\title{ \begin{flushright}
{\small CECS-PHY-06/04 \\  \vskip .4cm NSF-KITP-06-20\\
ULB-TH/06-05}
\end{flushright}
\vskip 1.0cm Asymptotic behavior and Hamiltonian analysis of anti-de Sitter
gravity coupled to scalar fields}
\author{Marc Henneaux$^{1,2}$, Cristi\'{a}n Mart\'{\i}nez$^{1}$, Ricardo
Troncoso$^{1}$, \\ Jorge Zanelli$^{1}$ \and $^{1}${\small Centro de Estudios
Cient\'{\i}ficos (CECS), Casilla 1469, Valdivia, Chile.} \and $^{2}${\small
Physique th\'{e}orique et math\'{e}matique, Universit\'{e} Libre de Bruxelles
and} \and {\small International Solvay Institutes, } \and{\small ULB-Campus
Plaine C.P.231, B-1050 Bruxelles, Belgium.}} \maketitle

\begin{abstract}
We examine anti-de Sitter gravity minimally coupled to a self-interacting
scalar field in $D\geq4$ dimensions when the
mass of the scalar field is in the range $m_{\ast}^{2}\leq m^{2}<m_{\ast}%
^{2}+l^{-2}$.  Here, $l$ is the AdS radius, and $m_{\ast}^{2}$\ is the
Breitenlohner-Freedman mass. We show that even though the scalar field
generically has a slow fall-off at infinity which back reacts on the metric so
as to modify its standard asymptotic behavior, one can still formulate
asymptotic conditions (i) that are anti-de Sitter invariant;  and (ii) that
allows the construction of well-defined and finite Hamiltonian generators for
all elements of the anti-de Sitter algebra.  This requires imposing a
functional relationship on the coefficients $a$, $b$ that control the two
independent terms in the asymptotic expansion of the scalar field. The anti-de
Sitter charges are found to involve a scalar field contribution. Subtleties
associated with the self-interactions of the scalar field as well as its
gravitational back reaction, not discussed in previous treatments, are
explicitly analyzed. In particular, it is shown that the fields develop extra
logarithmic branches for specific values of the scalar field mass (in addition
to the known logarithmic branch at the B-F bound).
\end{abstract}

\vskip 1.0cm

\section{Introduction}
\setcounter{equation}{0}

Anti-de Sitter gravity coupled to scalar fields with mass above the
Breitenlohner-Freedman bound \footnote{Here $l$ is the radius of
$D$-dimensional anti-de Sitter spacetime.} \cite{B-F,M-T}
\begin{equation}
m_{\ast}^{2}=-\frac{(D-1)^{2}}{4l^{2}},\label{BFbound}
\end{equation}
has generated considerable attention recently as it admits black hole solutions
with interesting new properties
\cite{HMTZ1,hairy,Hertog-Maeda1,MTZTop,Hertog-Maeda2,Gubser,3D}. The theory
supports solitons \cite{solitons} and provides a novel testing ground for
investigating the validity of the cosmic censorship conjecture
\cite{Controversy,HH}.   It is, of course, also relevant to the AdS/CFT
correspondence \cite{AdS-CFT}.

Boundary conditions in anti-de Sitter space are notoriously known to be a
subtle subject as information can leak out to or get in from spatial infinity
in a finite time.  Following \cite{B-F}, precise AdS asymptotic boundary
conditions on the metric were given in
\cite{Henneaux-Teitelboim,Henneaux-D,Brown-Henneaux} in the absence of matter
fields (or for localized matter).  It turns out, however, that these boundary
conditions do not accommodate generic scalar fields compatible with anti-de
Sitter symmetry when the mass $m$ of the scalar field is in the range
\begin{equation} m_{\ast}^{2}\leq m^{2}<m_{\ast}^{2}+\frac{1}{l^{2}} \ .
\label{Allowed-range}
\end{equation}

The main point can be already grasped by considering a free scalar field $\phi$
in  anti-de Sitter space,
\begin{equation} ds^2 = - \left(1 + \frac{r^2}{l^2} \right) dt^2 +
\left(1 + \frac{r^2}{l^2} \right)^{-1} dr^2 + r^2 d \Omega^2
\end{equation}
that behaves asymptotically as $\phi \sim r^{- \Delta}$ with $\Delta$ real. If
the exponent $\Delta$ is strictly greater than
\begin{equation} \Delta_R = \frac{D-3}{2} \ ,
\label{DeltaR}
\end{equation}
the scalar field is normalizable in the sense that the spatial integral of the
zeroth component $j^0$ of the Klein-Gordon current is finite. The condition
$\Delta > \Delta_R$ is thus necessary for the scalar field configuration to be
physically acceptable.  If, furthermore, the exponent is strictly greater than
\begin{equation} \Delta_{\ast} = \frac{D-1}{2} \ ,
\label{DeltaS}
\end{equation}
the Hamiltonian for the scalar field
\begin{equation} H= \frac{1}{2}\int d^{D-1}x \, \sqrt{g} \,
\sqrt{-g_{00}} \left(\left(\frac{\pi}{\sqrt{g}}\right)^2 + g^{ij}
\partial_i \phi \partial_j \phi \right)
\end{equation}
is a well-defined generator \cite{Regge-Teitelboim} as it stands and does not
need to be supplemented by a surface integral at infinity.  But if $\Delta \leq
\Delta_{\ast}$ (while remaining greater than the normalizability bound
$\Delta_R$), then the scalar field does contribute to surface integrals at
infinity and, when coupled to gravity, modifies the standard analysis of
asymptotically anti-de Sitter spaces.

Now, it follows from the Klein-Gordon equation that at large spatial distances,
neglecting the self-interaction, a scalar field of mass $m$, minimally coupled
to an AdS background, is asymptotically given by
\begin{equation}
\phi\sim\frac{a}{r^{\Delta_{-}}}+\frac{b}{r^{\Delta_{+}}}%
\ ,\label{Asympt-Phi-Linear}%
\end{equation}
where
\begin{equation}
\Delta_{\pm}=\frac{D-1}{2}\left(  1\pm\sqrt{1+\frac{4l^{2}m^{2}}{(D-1)^{2}}%
}\right)  \;,\label{Deltas}%
\end{equation}
are the roots of $\Delta(D-1-\Delta)+m^{2}l^{2}=0$. The coefficients $a$, $b$,
which depend on $t$ and the angles, are different from zero for generic
solutions. In particular, they do not vanish for the black hole or soliton
solutions considered in the literature.  Comparing (\ref{Deltas}) with
(\ref{DeltaR}) shows that in the range (\ref{Allowed-range}), both the
$a$-branch and the $b$-branch fulfill $\Delta_{\pm} > \Delta_R$ and are
physically acceptable. However, only the $b$-branch fulfills the stronger
condition $\Delta_+ > \Delta_{\ast}$ and hence does not contribute to surface
integrals at infinity (except when the B-F bound is saturated, in which case
$\Delta_+ = \Delta_{\ast}$\footnote{This limiting case where $\Delta_- =
\Delta_+$ needs a separate discussion as the asymptotic behavior of $\phi$
involves then also a logarithmic term.  That discussion was given in
\cite{HMTZ2} and is recalled in Section \ref{logarithmic} below.}). The
$a$-branch is always such that $\Delta_- \leq \Delta_{\ast}$ and does
contribute to surface integrals at infinity.

In the coupled Einstein-scalar system, taking into account the back reaction of
the scalar field on the geometry, one finds that the metric approaches anti-de
Sitter space at infinity more slowly when $a \not= 0$ than in the absence of
matter: the boundary conditions of \cite{Henneaux-Teitelboim} cannot
accommodate the $a$-branch and must be modified.  Because of this, the standard
surface term giving the energy in an asymptotically anti-de Sitter space
\cite{Abbott,Henneaux-Teitelboim} diverges. At the same time, there is a
further contribution to the surface integrals coming from the scalar field with
the possibility of cancelation of the divergences.

An additional effect arises when the mass reaches the value
\begin{equation} \label{m1}
m^2_{\ast} + \frac{(D-1)^2}{36 \,l^2},
\end{equation}
(which is in the allowed range if $D<7$).  Indeed, the asymptotic behavior
given in (\ref{Asympt-Phi-Linear}) is then changed, for generic potentials, by
a term of order $r^{-2 \Delta_-}\ln(r)$, which dominates over the $b$-branch
and is thus also non-negligible at infinity, contributing with further
divergencies to the surface integrals. When the mass exceeds the value
(\ref{m1}), the asymptotic behavior of the scalar field instead picks up a
relevant term of order $r^{-2 \Delta_-}$. Similarly, when the mass equals the
value
\begin{equation}\label{m2}
m^2_{\ast} + \frac{(D-1)^2}{16 \,l^2},
\end{equation} (which is in the allowed range if $D\leq4$), an additional term
of the form $r^{-3 \Delta_-}\ln(r)$ becomes relevant and also contributes to
the surface integrals. When the mass exceeds (\ref{m2}), the scalar field
acquires instead an extra term of order $r^{-3 \Delta_-}$ which is also
relevant. Finally, when the mass takes the value
\begin{equation} \label{m3}
m^2_{\ast} + \frac{9 (D-1)^2}{100 \,l^2}
\end{equation}
(which is also in the allowed range if $D\leq4$), a term $r^{-4
\Delta_-}\ln(r)$ becomes also relevant and must be taken into account, and for
larger mass  the scalar field possesses an extra term of the form $r^{-4
\Delta_-}$ instead of a logarithmic branch.

The purpose of this paper is to show that it is possible to relax the boundary
conditions on the scalar and gravitational fields in a way that allows for a
non-vanishing $a$-branch of the scalar field. These conditions are fully
compatible with asymptotic anti-de Sitter symmetry in the sense that they allow
for a consistent Hamiltonian formulation of the dynamics with well-defined,
finite, anti-de Sitter generators\footnote{To accommodate the $b$-branch alone
($a = 0$) presents no difficulty since the scalar field is then compatible with
the standard fall-off of the metric. Furthermore, it does not contribute to
surface integrals, except when the B-F bound is saturated, as recalled below.}.
With these boundary conditions, all divergences, including those arising from
the subleading terms $r^{-2 \Delta_-}$, $r^{-3 \Delta_-}$ and $r^{-4
\Delta_-}$, whenever relevant, consistently cancel. This also holds when the
logarithmic branches are present.

A notable feature of these boundary conditions is that they force $a$ and $b$
to be functionally related since otherwise the surface terms giving the
variations of the charges would not be integrable and hence the charges would
not exist. Hence, the functions $a$ and $b$ are found not to be independent.
The precise functional relationship between $a$ and $b$ is furthermore fixed by
anti-de Sitter symmetry (but is arbitrary if one demands only existence of the
surface integral defining the energy).

Our paper extends and completes previous work on the subject.

\begin{itemize}

\item The case of three spacetime dimensions was studied in
\cite{HMTZ1} for a particular value of the mass of the scalar field.  It was
already found there that a relationship must be imposed between $a$ and $b$ and
that the surface terms get scalar field contributions that cancel the
divergences.

\item The particular case when the B-F bound is saturated was
treated in all dimensions in \cite{HMTZ2}.  This case is peculiar on two
accounts: first, there is a logarithmic term in the expansion for the scalar
field because $\Delta_+ = \Delta_-$; second, both branches are relevant to the
surface integrals.

\item An analysis which turns out to be valid when the scalar mass
is in the range $m^2_{\ast} \leq m^2 < m^2_{\ast} + \frac{(D-1)^2}{36 \,l^2}$
was provided in all dimensions in \cite{Hertog-Maeda1}.
\end{itemize}

The plan of the paper is as follows: In the next section, the standard
asymptotic conditions for matter-free anti-de Sitter gravity and the form of
the charge generators are reviewed.  In Section \ref{logarithmic} the case of
gravity and minimally coupled scalar fields with a logarithmic branch is
summarized. A detailed analysis of the consequences of admitting both branches
in four dimensions is presented in Section \ref{fourD}.  In Section
\ref{HigherD}, the generalization for higher dimensions is discussed. Section
\ref{logs} contains the analysis of the special cases (\ref{m1}) when the
fields develop logarithmic branches. Section \ref{breaking} contains comments
concerning possible extensions of these results when the AdS symmetry is
broken. We conclude in section \ref{conclusions}.

We adopt the following conventions:  the action for gravity with a minimally
coupled self-interacting scalar field in $D\geq4$ dimensions is given by
\begin{equation}
I[g,\phi]=\int d^{D}x\sqrt{-g}\left(  \frac{R-2\Lambda}{2 \kappa}-\frac{1}%
{2}(\nabla\phi)^{2}-\frac{m^{2}}{2}\phi^{2}-U(\phi)\right)  \;, \label{action}
\end{equation}
where the self-interacting potential $U(\phi)$ is assumed to have an analytic
expansion in $\phi$ and is at least cubic in $\phi$. When unwritten, the
gravitational constant $\kappa=8\pi G$ is chosen as 1 and the cosmological
constant $\Lambda$ is $\Lambda =-l^{-2}(D-1)(D-2)/2$.

\section{Standard asymptotic conditions for gravity and charge generators}
\setcounter{equation}{0}

We first recall the standard situation, from which we shall depart in the
presence of a scalar field. In order to write down the asymptotic behavior of
the fields, the metric is written as $g_{\mu\nu}=\bar{g}_{\mu\nu}+h_{\mu\nu}$,
where $h_{\mu\nu}$ is the deviation from the AdS metric,
\begin{equation}
d\bar{s}^{2}=-(1+r^{2}/l^{2})dt^{2}+(1+r^{2}/l^{2})^{-1}dr^{2}+r^{2}%
d\Omega^{D-2}\;.
\end{equation}
For matter-free gravity, the asymptotic behavior of the metric is given in
\cite{Henneaux-Teitelboim,Henneaux-D,Brown-Henneaux} and reads
\begin{equation}%
\begin{array}
[c]{lll}%
h_{rr} & = & \displaystyle O(r^{-D-1})\;,\\[2mm]%
h_{rm} & = & O(r^{-D})\;,\\[1mm]%
h_{mn} & = & O(r^{-D+3}).\;
\end{array}
\label{Standard-Asympt}%
\end{equation}
Here the indices have been split as $\mu=(r,m)$, where $m$ includes the time
coordinate $t$ and the $D-2$ angles.

The asymptotic symmetries correspond to the diffeomorphisms that map the
asymptotic conditions into themselves, i. e. , $\xi^{\mu}$ generates an
asymptotic symmetry if
\[
\mathcal{L}_{\xi}g_{\mu\nu}=O(h_{\mu\nu})\ .
\]
Note that it is not necessary to require the existence of exact Killing vectors
when dealing with conserved charges for a generic configuration in gravity,
\emph{e.g.} for the dynamics of several objects. This can be seen intuitively,
since in a region far from the objects only the leading terms are relevant to
compute the energy and thus only the existence of an asymptotic timelike
Killing vector is required. Analogously, the linear and angular momenta can be
obtained through the asymptotic symmetries.

It is easy to check that the asymptotic conditions (\ref{Standard-Asympt}) are
invariant under $SO(D-1,2)$ for $D\geq 4$, and under the infinite-dimensional
conformal group in two dimensions (two copies of the Virasoro algebra) for
$D=3$. The asymptotic behavior of a generic asymptotic Killing vector field
$\xi^{\mu}$ is given by
\begin{equation}%
\begin{array}
[c]{lllllll}%
\xi^{r} & = & O(r), &  & \xi_{,r}^{r} & = & O(1)\\
\xi^{m} & = & O(1), &  & \xi_{,r}^{m} & = & O(r^{-3})
\end{array}
\label{Asympt-Symm}%
\end{equation}

The charges that generate the asymptotic symmetries involve only the metric and
its derivatives, and are given by
\begin{equation}
Q_{0}(\xi)=\int d^{D-2}S_{i}\left(   \frac{\bar{G}^{ijkl}}{2\kappa}(\xi^{\bot
}h_{kl|j}-\xi_{,j}^{\bot}h_{kl})+2\xi^{j} \pi_{j}^{\;\;i}\right)
\;,\label{Q0}%
\end{equation}
where the supermetric is defined as
$G^{ijkl}=\frac{1}{2}g^{1/2}(g^{ik}g^{jl}+g^{il}g^{jk}-2g^{ij}g^{kl})$, and the
vertical bar denotes covariant differentiation with respect to the spatial AdS
background. From (\ref{Standard-Asympt}) it follows that the momenta possess
the following fall-off at infinity
\begin{equation}
\pi^{rr}=O(r^{-1}),\quad\pi^{rm}=O(r^{-2}),\quad\pi^{mn}=O(r^{-5})\;,
\end{equation}
and hence, the surface integral (\ref{Q0}) is finite.  We have adjusted the
constants of integration in the charges so that anti-de Sitter space has zero
anti-de Sitter charges.

The Poisson bracket algebra of the charges yields the AdS group for $D>3$ and
two copies of the Virasoro algebra with a central charge given by
\begin{equation}
c=\frac{3l}{2G},
\label{Central Charge}%
\end{equation}
in three dimensions \cite{Brown-Henneaux}.

The asymptotic conditions (\ref{Standard-Asympt}) hold not only in the absence
of matter but also for localized matter fields which fall off sufficiently fast
at infinity, so as to give no contributions to the surface integrals defining
the generators of the asymptotic symmetries. Note that, as pointed out above, a
minimally coupled scalar field would not contribute to the charges if it goes
as $\phi\sim r^{-((D-1)/2+\varepsilon)}$ for large $r$.

\section{Gravity and scalar fields saturating the \\
Breitenlohner-Freedman bound} \label{logarithmic} \setcounter{equation}{0}

When the scalar field mass saturates the Breitenlohner--Freedman bound, i. e.,
for $m^{2}=m_{\ast}^{2}$,  the scalar field acquires a logarithmic fall-off.
This induces a back-reaction on the metric which differs from the standard
asymptotic behavior by the addition of logarithmic terms as well. This case was
treated in \cite{HMTZ2} and we recall the results here for completeness. The
leading terms for $h_{\mu\nu}$ and $\phi$ as $r\rightarrow\infty$ are found to
be\footnote{Here the roles of $a$ and $b$ are interchanged with respect to
those in Ref.\cite{HMTZ2}.}
\begin{align}
&
\begin{array}
[c]{lll}%
\phi & = & \!\displaystyle\frac{1}{r^{(D-1)/2}}\left(a\ln \left(
r/r_{0}\right)  +b \right)  +\cdot\cdot\cdot\\[3mm]
&  &
\end{array}
\nonumber\\
&
\begin{array}
[c]{lll}%
h_{rr} & = & \!\displaystyle-\frac{(D-1)l^{2}a^{2}}{2(D-2)}\;\frac{\ln
^{2}\left(  r/r_{0}\right)  }{r^{(D+1)}}\\[3mm]
& + & \!\displaystyle\frac{l^{2}(a^{2}-(D-1)ab)}{D-2}\;\frac{\ln \left(
r/r_{0}\right)  }{r^{(D+1)}}+O\left(  \frac{1}{r^{(D+1)}}\right) \\[3mm]%
h_{mn} & = & \!\displaystyle O\left(  \frac{1}{r^{(D-3)}}\right) \\[3mm]%
h_{mr} & = & \!\displaystyle O\left(  \frac{1}{r^{(D-2)}}\right) \\[3mm]
&  &
\end{array}
\end{align}
where $a=a(x^{m})$, $b=b(x^{m})$, and $r_{0}$ is an arbitrary
constant\footnote{{}
Making use of the the relaxed asymptotic conditions, the momenta at infinity
are found to be
\begin{align}
\pi^{rr}=O(r^{-1}),\quad &  \pi^{rm}=O(r^{-2}),\quad\pi^{mn}=O(r^{-5}\ln
^{2}(r)),\\
&  \pi_{\phi}=O(r^{(D-7)/2}\ln(r))\,,
\end{align}
}. This relaxed asymptotic behavior still preserves the original asymptotic
symmetry, which is $SO(D-1,2)$ for $D\geq4$, and the conformal group in two
spacetime dimensions for $D=3$, as indicated in the previous section.

The variation of the corresponding conserved charges can be obtained following
the Regge-Teitelboim approach \cite{Regge-Teitelboim} and it is found to depend
on $\delta a$ and $\delta b$. This differential is exact --and hence,
integration of the variation of these charges as local functionals of the
fields  is possible-- only is $a$ and $b$ are functionally related. AdS
invariance fixes the relation to be of the form
\begin{equation} \label{ablog}
b=-\frac{2}{(D-1)}\,a\,\ln(a/a_{0})\; ,
\end{equation} where $a_{0}$ is a
constant.

The conserved charges acquire an extra contribution coming from the scalar
field which read%
\begin{equation}
Q(\xi)=Q_{0}(\xi)+Q_{\phi}\ ,\label{Qtotal}%
\end{equation}
where $Q_{0}(\xi)$ is given by (\ref{Q0}), and $Q_{\phi}$ is given by%
\begin{equation} \label{Qphilog}
Q_{\phi}=\frac{1}{2(D-1)l}\int d\Omega^{D-2}r^{(D-2)}\xi^{\bot}\left\{
l^2(n^r\partial_r\phi)^{2}+\frac{\left(  D-1\right)  ^{2}}{4}\phi ^{2}\right\}
\end{equation}
where $n^r=(\sqrt{g_{rr}})^{-1}$ is the only nonvanishing component of the unit
normal to the boundary. Note that here, one can replace $g_{rr}$ by the
background value $\bar{g}_{rr}$. These charges are finite even when the
logarithmic branch is switched on, because the divergence in the gravitational
piece is canceled by the divergence in the scalar piece.

In the case $a=0$, the asymptotic behavior of the metric reduces to the
standard one (\ref{Standard-Asympt}), and the original asymptotic symmetry is
preserved. Nevertheless, the charges (\ref{Qtotal}) still give a non-trivial
contribution coming from the scalar field since the exponent is just equal to
$\Delta_{*}$. The algebra of the charges (\ref{Qtotal}) is identical to the
standard one discussed in the previous section.

In the absence of the logarithmic branch, conserved charges have been
constructed in \cite{Hollands-Ishibashi-Marolf} following covariant methods,
and a comparison of different methods to compute the mass of five-dimensional
rotating black holes in supergravity has been performed in \cite{Chen-Lu-Pope}%
. We note that the AdS charges of metric-scalar field configurations with a
logarithmic branch could also be computed through the method of holographic
renormalization, as in \cite{BFS1,BFS2} in five dimensions.

\section{Asymptotically AdS gravity with a minimally coupled scalar field
- The case of four dimensions} \label{fourD} \setcounter{equation}{0}

We now turn to the case where the scalar field mass is strictly above the
Breitenlohner--Freedman bound. We treat $D=4$ first, as it is for this
spacetime dimension that the non-linearities due to the potential and to the
interactions with the gravitational field are the most intricate.  Matters
simplify in higher dimensions. We expand the potential up to the relevant
order,
\[
l^{2}U=C_{3}\phi^{3}+C_{4}\phi^{4}+C_{5}\phi^{5}+O(\phi^{6})\ .
\]
As explained in the introduction, there are four cases in which the fields have
a power-law decay:
\begin{itemize} \item $m^2_{\ast}  < m^2 < m^2_{\ast} +\frac{1}{4 \, l^2}$
\item $m^2_{\ast} + \frac{1}{4 \, l^2} < m^2 < m^2_{\ast} +\frac{9}{16 \, l^2}$
\item $m^2_{\ast} + \frac{9}{16 \, l^2} < m^2 < m^2_{\ast} +\frac{81}{100 \, l^2}$
\item $m^2_{\ast} + \frac{81}{100 \,l^2} < m^2< m^2_{\ast} +\frac{1}{l^2}$
\end{itemize}
For the limiting cases $m^2= m_*^2+(2l)^{-2}$, $m^2_{\ast} + \frac{9}{16 \,
l^2}$, and $m^2_{\ast} + \frac{81}{100 \,l^2}$, the fields acquire logarithmic
branches, as discussed in Sect. \ref{logs}.

We first give the asymptotic conditions. We explain next how they were arrived
at. We start with the last range, which displays the full complexity of the
problem.

\subsection{$m^2_{\ast} + \frac{81}{100 \,l^2} <m^2< m^2_{\ast} +\frac{1}{ l^2}$}

In this range, the exponent lies in the range $1/2<\Delta_-<3/5$, while
$\Delta_+$ varies between $12/5$ and  $5/2$. It follows that $r^{-\Delta_-}$,
$r^{-2\Delta_-}$, $r^{-3\Delta_-}$ and $r^{-4 \Delta_-}$ (but not
$r^{-5\Delta_-}$) dominate asymptotically $r^{- \Delta_+}$. We shall denote
from now on $\Delta_-$ simply by $\Delta$.

The appropriate asymptotic conditions for the scalar field and the metric are
given by \begin{align} \phi &
=ar^{-\Delta}+\beta_{1}a^{2}r^{-2\Delta}+\beta_{2}a^{3}r^{-3\Delta }
+\beta_{3}a^{4}r^{-4\Delta}+br^{-\Delta_+}+\cdot\cdot\cdot
\ .\label{AsymptPhi4}\\
& \nonumber
\end{align}%
\begin{equation}%
\begin{array}
[c]{lll}%
h_{rr} & = & \displaystyle\frac{\kappa l^{2}}{r^{2}}\left(  \alpha_{1}a^{2}%
r^{-2\Delta}+\alpha_{2}a^{3}r^{-3\Delta}+ \alpha_{3}a^{4}r^{-4\Delta}+
\alpha_{4} a^{5}r^{-5\Delta}\right) \\
& + & \displaystyle\frac{f_{rr}}{r^{5}}+\cdots\\[3mm]%
h_{mn} & = & \displaystyle\frac{f_{mn}}{r}+\cdots\\[3mm]%
h_{mr} & = & \displaystyle\frac{f_{mr}}{r^{2}}+\cdots
\end{array}
\label{AsymptMetric4}%
\end{equation}
where the dots $(\cdot\cdot\cdot)$ indicate subleading terms that do not
contribute to the charges. Here, $b$, $f_{rr}$ $f_{mn}$ and $f_{rm}$ are
independent functions of time and the angles ($x^{m}$).  The function $a$ is
determined by $b$ through
\begin{equation}
a=a_{0}b^{\frac{\Delta}{\Delta_{+}}}\ ,\label{ab4AdS}%
\end{equation}
where $a_{0}$ is an arbitrary constant. The standard asymptotic conditions
($a$-branch switched off) are recovered for $a_{0}=0$, while $a_0 = \infty$
corresponds to switching off the $b$-branch. The coefficients $\beta_{1}$,
$\beta_{2}$, $\beta_{3}$ are constants given by the following expressions:
\begin{align}
\beta_{1}  &  =\bar{\beta}_{1}\ ,\nonumber\\
\beta_{2}  &  =\bar{\beta}_{2}+\kappa\frac{ \Delta(3-2 \Delta)}{4(4\Delta
-3)}\ ,\label{Betas-Gravt}\\
\beta_{3}  &  =\bar{\beta}_{3}+\kappa\frac{ C_{3}(-153+327\Delta-170\Delta^{2}%
)}{18(\Delta-1)(4\Delta-3)(5\Delta-3)}\ ,\nonumber
\end{align}
where the coefficients $\bar{\beta}_1$, $\bar{\beta}_2$ and $\bar{\beta}_3$
correspond to those found neglecting the back reaction, i. e., for a fixed AdS
background,
\begin{align}
\bar{\beta}_{1}  &  =\frac{C_{3}}{\Delta(\Delta-1)},\nonumber\\
\bar{\beta}_{2}  &  =\frac{2C_{4}}{\Delta(4\Delta-3)}+\frac{3C_{3}^{2}}%
{\Delta^{2}(\Delta-1)(4\Delta-3)},\nonumber\\
\bar{\beta}_{3}  &  =\frac{5C_{5}}{3\Delta(5\Delta-3)}+\frac{4C_{3}%
C_{4}(5\Delta-4)}{\Delta^{2}(\Delta-1)(4\Delta-3)(5\Delta-3)}%
+\label{Betas-AdS}\\
&  \;+\frac{C_{3}^{3}(10\Delta-9)}{\Delta^{3}(5\Delta-3)(4\Delta
-3)(\Delta-1)^{2}}\ .\nonumber
\end{align}
Similarly, the constants $\alpha_1$, ..., $\alpha_4$ in the metric are given by
\begin{align}
\alpha_{1}  &  =-\frac{\Delta}{2}\ ,\nonumber\\
\alpha_{2}  &  =-\frac{4}{3}\Delta\beta_{1}\ ,\nonumber\\
\alpha_{3}  &  =-\frac{\Delta}{4}\left( -\frac{\kappa\Delta}{2}+6\beta_{2}%
+4\beta_{1}^{2}\right)  \ ,\label{alphas-charges-4}\\
\alpha_{4}  &  =-\frac{\Delta}{5}\left(  8\beta_{3}+12\beta_{1}\beta_{2}%
-\frac{10}{3}\kappa\Delta\beta_{1}\right)  \ .\nonumber
\end{align}

Note that when the $a$-branch of the scalar field is switched on, the
asymptotic fall-off of the metric acquire a strong back reaction in comparison
with the standard asymptotic conditions in Eq. (\ref{Standard-Asympt}). In
fact, the first two terms in $h^{rr}$ dominate asymptotically the $1$ in the
$g^{rr}$-component of the anti-de Sitter background metric. In turn, the
effects of the gravitational back reaction as well as of the potential
drastically modify the asymptotic behavior of the scalar field, as can be seen
by comparing Eq. (\ref{AsymptPhi4}) with the behavior obtained for the linear
approximation in a fixed AdS background in Eq. (\ref{Asympt-Phi-Linear}), where
the asymptotically relevant powers $r^{-2\Delta}$, $r^{-3 \Delta}$ and $r^{-4
\Delta}$ are absent.  Note also that the effects of the self interactions are
relevant even if the gravitational field is switched-off.

It is interesting to point out that the back reaction of the gravitational
field has a similar effect on the asymptotic form of the scalar field as the
presence of cubic and quartic self-interaction terms, since even in the absence
of a self-interacting potential, the term $\beta _{2}a^{3}$ must be considered.

\subsection{$m^2_{\ast} + \frac{9}{16 \, l^2}
< m^2 < m^2_{\ast} +\frac{81}{100 \, l^2}$}

In this case, the exponent $\Delta$ is in the range $3/5<\Delta<3/4$, while
$\Delta_+$ varies between $9/4$ and $12/5$. It follows that $r^{-\Delta}$,
$r^{-2\Delta}$ and $r^{-3\Delta}$ (but not $r^{-4\Delta}$) dominate
asymptotically over $r^{- \Delta_+}$.

The appropriate asymptotic conditions for the scalar field and the metric are
given by
\begin{align} \phi &
=ar^{-\Delta}+\beta_{1}a^{2}r^{-2\Delta}+\beta_{2}a^{3}r^{-3\Delta
}+br^{-(\Delta+\gamma)}+\cdot\cdot\cdot
\ .\label{AsymptPhi4b}\\
& \nonumber
\end{align}%
\begin{equation}%
\begin{array}
[c]{lll}%
h_{rr} & = & \displaystyle\frac{\kappa l^{2}}{r^{2}}\left(  \alpha_{1}a^{2}%
r^{-2\Delta}+\alpha_{2}a^{3}r^{-3\Delta}+ \alpha_{3} a^{4}r^{-4\Delta}\right) \\
& + & \displaystyle\frac{f_{rr}}{r^{5}}+\cdots\\[3mm]%
h_{mn} & = & \displaystyle\frac{f_{mn}}{r}+\cdots\\[3mm]%
h_{mr} & = & \displaystyle\frac{f_{mr}}{r^{2}}+\cdots
\end{array}
\label{AsymptMetric4b}%
\end{equation}
with $a$ related to $b$ as in (\ref{ab4AdS}) and $\beta_1$, $\beta_2$,
$\alpha_1$, $\alpha_2$ and $\alpha_3$ given by (\ref{Betas-Gravt}),
(\ref{Betas-AdS}) and (\ref{alphas-charges-4}).  Note that the coefficient
$C_5$ of the potential does not enter the relevant expressions and hence, its
precise value need not be specified.

\subsection{$m^2_{\ast} + \frac{1}{4 \, l^2}
< m^2 < m^2_{\ast} +\frac{9}{16 \, l^2}$}

In this range, the exponent $\Delta$ varies between $3/4$ and $1$, while
$\Delta_+$ varies between $2$ and  $9/4$. It follows that $r^{-\Delta}$ and
$r^{-2\Delta}$  (but not $r^{-3\Delta}$) dominate asymptotically $r^{-
\Delta_+}$.

The appropriate asymptotic conditions for the scalar field and the metric are
given by
\begin{equation}%
\begin{array}
[c]{lll}%
\phi & = & ar^{-\Delta}+\beta_{1}a^{2}r^{-2\Delta}+br^{-\Delta_+}+\cdots \\
h_{rr} & = & \displaystyle\frac{\kappa l^{2}}{r^{2}}\left(  \alpha_{1}a^{2}%
r^{-2\Delta}+\alpha_{2}a^{3}r^{-3\Delta}\right) +
\displaystyle\frac{f_{rr}}{r^{5}}+\cdots\\[3mm]%
h_{mn} & = & \displaystyle\frac{f_{mn}}{r}+\cdots\\[3mm]%
h_{mr} & = & \displaystyle\frac{f_{mr}}{r^{2}}+\cdots
\end{array}
\label{AsymptMetric4c}%
\end{equation}
with $a$ related to $b$ as in (\ref{ab4AdS}) and $\beta_1$, $\alpha_1$ and
$\alpha_2$ given by (\ref{Betas-Gravt}), (\ref{Betas-AdS}) and
(\ref{alphas-charges-4}). Note that now it is not necessary to specify the
coefficients $C_4$ and $C_5$, since they do not enter the relevant expressions.

\subsection{$m^2_{\ast}  < m^2 < m^2_{\ast} +\frac{1}{4 \, l^2}$}

The range of the exponent $\Delta$ is now $1<\Delta< 3/2$, while $\Delta_+$
varies between $3/2$ and $2$. It follows that $r^{-\Delta}$  (but not
$r^{-2\Delta}$) dominate asymptotically over $r^{- \Delta_+}$.

The appropriate asymptotic conditions for the scalar field and the metric are
given by
\begin{equation}%
\begin{array}
[c]{lll}%
\phi & = & ar^{-\Delta}+br^{-\Delta_+}+\cdots \\
h_{rr} & = & \displaystyle\frac{\kappa l^{2}}{r^{2}}\left( \alpha_{1}a^{2}%
r^{-2\Delta}\right)
 +  \displaystyle\frac{f_{rr}}{r^{5}}+\cdots\\[3mm]%
h_{mn} & = & \displaystyle\frac{f_{mn}}{r}+\cdots\\[3mm]%
h_{mr} & = & \displaystyle\frac{f_{mr}}{r^{2}}+\cdots
\end{array}
\label{AsymptMetric4d}%
\end{equation}
with $a$ related to $b$ as in (\ref{ab4AdS}) and  $\alpha_1$ given by
(\ref{alphas-charges-4}).  Note that in this range it is no longer necessary to
specify the potential.

We shall now justify the boundary conditions and check their consistency.
First, we verify they anti-de Sitter invariance. Second, we shall derive the
anti-de Sitter charges and show that all divergences cancel.  To carry the
analysis, we shall assume to begin with that the functions $a$ and $b$, as well
as the constants $\beta_i$ and $\alpha_i$ are arbitrary. The necessity to
restrict them as in the above formulas will then appear quite clearly.

\subsection{Asymptotic AdS symmetry}

It is easy to verify that, even in the presence of the extra terms in the
scalar field and in the metric, the asymptotic conditions given by Eqs.
(\ref{AsymptPhi4}) and (\ref{AsymptMetric4}) are preserved under the asymptotic
AdS symmetry. Indeed, since the action of an asymptotic Killing vector
$\xi^{\mu}=(\xi^{r},\xi^{m})$ on the scalar field reads,
\begin{equation}
\phi\rightarrow\phi+\mathcal{L}_{\xi}\phi=\phi+\xi^{\mu}\partial_{\mu}%
\phi\ ,\label{phi tilde}%
\end{equation}
where $\xi^{r}=\eta^{r}(t,x^{m})r+O(r^{-1})$, and $\xi^{m}=O(1)$, the
asymptotic behavior is of the same form as in Eq.(\ref{AsymptPhi4}) with
\begin{align}
a  &  \rightarrow a-\eta^{r}\Delta a+\xi^{m}\partial_{m}a\label{a tilde}\\
b  &  \rightarrow b-\eta^{r}\Delta_{+}b+\xi^{m}\partial_{m}b\ ,\label{b tilde}%
\end{align}
verifying that the asymptotic symmetries are preserved. Similarly, the Lie
derivative of the metric under the anti-de Sitter Killing vectors has the
requested fall-off. (Equations (\ref{a tilde}) and (\ref{b tilde}) are
generically modified if $\Delta_+/\Delta$ is an integer, since then there
appear logarithmic branches, as shown in Section \ref{logs}.)

However, as discussed below (and noticed in \cite{HMTZ1,HMTZ2,Hertog-Maeda1}),
the integration of the variation of the symmetry generators as local
functionals of the fields requires $a$ and $b$ to be functionally related in
the form $a=a(b,x^{m})$. Consistency of this assumption with the asymptotic AdS
symmetry requires the compatibility of Eqs. (\ref{a tilde}) and (\ref{b
tilde}), which means that
\begin{equation}
\eta^{r}\left(  \Delta a-\Delta_{+}b\frac{\partial a}{\partial b}\right)
+\xi^{m} \left(\frac{\partial a}{\partial b}\partial_{m} b-
\partial_{m}a\right)=0\ .\label{a(b)}
\end{equation}
Hence, since $\eta^{r}$ and $\xi^{m}$ are independent, the asymptotic AdS
symmetry fixes the functional relationship between $a$ and $b$ to be of the
form (\ref{ab4AdS}) given above.

\subsection{Anti-de Sitter Charges}

In order to write down the conserved charges, it is convenient to split the deviation $h_{ij}$ from the AdS background as
\begin{equation}
h_{ij}=\varphi_{ij}+\psi_{ij},
\end{equation}
where
\begin{equation}
\psi_{rr}  =  \displaystyle\frac{f_{rr}}{r^{5}}+O(r^{-6}), \quad
\psi_{mn}  =  \displaystyle\frac{f_{mn}}{r}+O(r^{-2}), \quad
\psi_{mr} =  \displaystyle\frac{f_{mr}}{r^{2}}+O(r^{-3}),
\end{equation}
and $\varphi_{ij}=h_{ij}-\psi_{ij}$. The $\psi_{ij}$-part contributes to finite surface integrals at infinity, while $\varphi_{ij}$, which collects the terms that go more slowly to zero, yields divergences.

The variation of the conserved charges corresponding to the asymptotic
symmetries can be found following the Regge-Teitelboim approach
\cite{Regge-Teitelboim}. We shall carry out the computation in the more complex
case $m^2_{\ast} + \frac{81}{100 \,l^2} < m^2< m^2_{\ast} +\frac{1}{ l^2}$ and
comment later on for the other ranges of the mass.

The contributions coming from gravity and the scalar field to the conserved
charges, $Q_{G}(\xi)$ and $Q_{\phi}(\xi)$ are respectively given by
\begin{align}
\delta Q_{G}(\xi)  &  =\frac{1}{2\kappa}\int d^{2}S_{l}\left[
G^{ijkl}(\xi^{\bot
}\delta g_{ij;k}-\xi_{,k}^{\bot}\delta g_{ij})\right. \label{DQGgen}\\
&  +\left.  \int d^{2}S_{l}(2\xi_{k}\delta\pi^{kl}+(2\xi^{k}%
\pi^{jl}-\xi^{l}\pi^{jk})\delta g_{jk})\right] \nonumber\\
\delta Q_{\phi}(\xi)  &  =-\int d^{2}S_{l}\left(  \xi^{\bot}g^{1/2}%
g^{lj}\partial_{j}\phi\delta\phi+\xi^{l}\pi_{\phi}\delta\phi\right)
\;.\label{DQPhigen}%
\end{align}

Using the relaxed asymptotic conditions (\ref{AsymptPhi4}),
(\ref{AsymptMetric4}), the momenta at infinity are found to be
\begin{align}
\pi^{rr}=O(r),\quad &
\pi^{rm}=O(r^{-2}),\quad\pi^{mn}=O(r^{-(2+\Delta)}),\label{PiG}\\
&  \pi_{\phi}=O(r^{-\Delta})\,,\label{PiPhi}%
\end{align}
(here the indices $m,n$ are purely angular) and hence Eq. (\ref{DQGgen}) acquires the form
\begin{align}
\delta Q_{G}(\xi)  &  =\delta Q_{G}(\xi)|_{\textrm{finite}}+\int d^{2}\Omega\frac
{\xi^{t}}{l^{2}}\left(  2\alpha_{1}a\delta
ar^{3-2\Delta}+3\alpha_{2}a^{2}\delta
ar^{3(1-\Delta)}\right. \\
&  \left.  +a^{3}\delta ar^{3-4\Delta}[4\alpha_{3}-3\kappa\alpha_{1}^{2}]+a^{4}\delta
ar^{3-5\Delta}\left(5\alpha_{4}-\frac{15}{2}\kappa\alpha_{1}\alpha_{2}\right) \right)
\ ,\nonumber
\end{align}
where $\delta Q_{G}(\xi)|_{\textrm{finite}}$ stands for the terms of
$O(1)$ and is given explicitly by \footnote{In the presence of the
scalar field, the terms of order $r^{-2}$ in $h_{mr}$ give a
nontrivial finite contribution to the charges. This is in contrast
with the standard case, where these terms can be gauged away
\cite{Henneaux-Teitelboim}.}
\begin{eqnarray}
\delta Q_{G}(\xi)|_{\textrm{finite}}  &  = &\frac{1}{2\kappa}\int d^{2}S_{l}
\bar{G}^{ijkl}(\xi^{\bot
}\delta \psi_{ij|k}-\xi_{,k}^{\bot}\delta \psi_{ij}) \label{DQGfin}\\
&  +& 2 \int d^{2}S_{l}\xi_{k}\delta\pi^{kl} . \nonumber
\end{eqnarray}

In a similar way, Eq. (\ref{DQPhigen}) takes the form
\begin{align}
\delta Q_{\phi}(\xi)  &  =\frac{\Delta}{l^{2}}\int d^{2}\Omega\xi^{t}\left(
a\delta ar^{3-2\Delta}+4\beta_{1}a^{2}\delta ar^{3(1-\Delta)}+a^{3}\delta
ar^{3-4\Delta}\left[  6\beta_{2}+4\beta_{1}^{2}-\kappa\frac{\alpha_{1}}{2}\right]
\right. \nonumber\\
&  \left.  +a^{4}\delta ar^{3-5\Delta}\left[  8\beta_{3}+12\beta_{1}%
\beta_{2}-2\kappa\alpha_{1}\beta_{1}-\kappa\frac{\alpha_{2}}{2}\right]  +a\delta b+b\delta
a\frac{\Delta_+}{\Delta}\right)  \ .\label{DeltaQPhiFinal4}%
\end{align}

Therefore, requiring the total variation of the charges, $\delta
Q=\delta Q_{G}+\delta Q_{\phi}$, to be finite forces the
coefficients $\alpha_1,\cdots, \alpha_4$ appearing in the asymptotic
form of the metric (\ref{AsymptMetric4}) to be fixed in terms of the
scalar field mass parameter and the $\beta^{\prime}s$ appearing in
the asymptotic form of the scalar field (\ref{AsymptPhi4}) exactly
as in (\ref{alphas-charges-4}). This is the rationale behind these
equations. Thus,  the variation of the charges becomes
\begin{equation} \label{vartotal}
\delta Q(\xi)= \delta Q_{G}(\xi)|_{\textrm{finite}}+\int d^{2}\Omega\frac{\xi^{t}}{l^{2}}\left(a\delta b \Delta+b\delta
a \Delta_+\right).
\end{equation}

Once the variation of the charges are guaranteed to be finite, one can ask the
question of whether they are integrable.  It is here that a functional
relationship on $a$ and $b$ must be imposed. Indeed, since $\Delta_+ /\Delta
\not= 1$,  the integrability of the variation of the matter piece of the
charges given by (\ref{DeltaQPhiFinal4}) and by (\ref{vartotal}) as a local functional of the fields
requires $\delta a$ and $\delta b$ not to be independent, i.e., $a$ and $b$
must be functionally related. As discussed above, the form (\ref{ab4AdS}) is
then forced by asymptotic AdS symmetry.

Now, we will integrate $ \delta Q_{G}(\xi)$ and $ \delta Q_{\phi}(\xi)$ separately as functions of the canonical variables. For matter piece, we get
\begin{equation}
Q_{\phi}(\xi)=\frac{1}{6l}\int d^{2}\Omega r^{2}\xi^{\perp}\left[
l^2(n^r\partial_r\phi)^{2} -m^2 l^2 \phi^{2}+k_{3}\phi^{3}+k_{4}\phi^{4}
+k_{5}\phi^{5}\right]  \ ,\label{QPhi4}
\end{equation}
with
\begin{eqnarray}
k_3 & = & -2C_3 \nonumber\\
k_4 & = & -2C_4 - \kappa \frac{3}{8}\Delta^2   \label{ks}\\
k_5 & = & -2C_5 - \kappa \frac{C_3 \Delta}{2(\Delta-1)} \nonumber
\end{eqnarray}
Note that the gravitational correction to $k_4$ does not depend on the
potential, whereas the gravitational correction to $k_5$ is proportional to the
coupling constant of $\phi^3$.

The normal is given as before by $n^r=(\sqrt{g_{rr}})^{-1}$ but now one cannot
replace it by its background value $(\sqrt{\bar{g}_{rr}})^{-1}$ when $m^2 \geq
m^2_{\ast} + \frac{9}{16 \, l^2}$.

Similarly, the purely gravitational part of the charge can also be integrated
to yield
\begin{equation}
Q_{G}(\xi)=Q_{0}(\xi)+\Delta Q(\xi)\;,\label{Q g}%
\end{equation}
where $Q_{0}$ is given by the standard formula in Eq. (\ref{Q0}), and
\begin{equation}
\Delta Q(\xi)=-\frac{3}{4\kappa}\int_{\partial\Sigma}d^{2}\Omega\frac{r^{6}}{l^{5}%
}\xi^{\perp}h_{rr}^{2}\;\label{DeltaQ}%
\end{equation}
is a nonlinear correction in the deviation from the background metric that
arises because one cannot replace the  supermetric $G^{ijkl}$  by its
background value at infinity: the difference does contribute to the surface
integral. This nonlinear term in the deviation $h_{rr}$ could not have been
obtained through standard perturbative methods to construct conserved charges.
and is essential to make the charges finite (it cancels some divergences). A
similar phenomenon was observed in \cite{HMTZ1} in 2+1 dimensions and in
\cite{BarCom} in the context of Goedel black holes.

The symmetry generators are then finite and given by%
\[
Q(\xi)=Q_{G}(\xi)+Q_{\phi}(\xi)\ ,
\]
with $Q_{G}(\xi)$ and $Q_{\phi}(\xi)$ given by Eqs. (\ref{Q g}) and
(\ref{QPhi4}), respectively.

An expression for $Q(\xi)$ which is manifestly free of divergences is easily obtained by inserting the asymptotic expressions of the fields and using the relationship between $a$ and $b$, and reads
\begin{equation} \label{qtotal}
\displaystyle Q(\xi)= Q_{0}(\xi)|_{\textrm{finite}}-\frac{2}{3} m^2 a_0^{-\frac{\Delta_+}{\Delta}}\int d^{2}\Omega \xi^{t} a^{\frac{3}{\Delta}} ,
\end{equation}
where
\begin{equation} \label{qofin}
Q_{0}(\xi)|_{\textrm{finite}}= \int d^{2}S_{i}\left(   \frac{\bar{G}^{ijkl}}{2\kappa}(\xi^{\bot
}\psi_{kl|j}-\xi_{,j}^{\bot}\psi_{kl})+2\xi^{j} \pi_{j}^{\;\;i}\right).
\end{equation}
The last term in Eq. (\ref{qtotal}) can be written in terms of $\phi$. Then, we obtain
\begin{equation} \label{qtotal2}
\displaystyle Q(\xi)= Q_{0}(\xi)|_{\textrm{finite}}-\frac{2}{3} m^2 a_0^{-\frac{\Delta_+}{\Delta}}\int d^{2}\Omega \xi^{t} r^3 \phi^{\frac{3}{\Delta}} .
\end{equation}
In the case of spherical symmetry, the energy ($\xi=\frac{\partial}{\partial t}$) is given by
\begin{equation} \label{qtotal3}
\displaystyle Q(\frac{\partial}{\partial t})= \frac{ 4 \pi f_{rr}}{\kappa l^4}-\frac{8 \pi}{3} m^2 a_0^{-\frac{\Delta_+}{\Delta}} a^{\frac{3}{\Delta}} .
\end{equation}

It should be stressed that only the sum of the
gravitational contribution \textit{and} the one of scalar field defines a
meaningful AdS charge that is conserved. Each term separately may vary as one
makes asymptotic AdS time translations. The algebra of the charges
(\ref{Qtotal}) is identical to the standard one, i. e., the AdS algebra. This
can be readily obtained following Ref \cite{Brown-Henneaux2}, where it is shown
that the bracket of two charges provides a realization of the asymptotic
symmetry algebra with a possible central extension.

For the other ranges of the mass, the analysis proceeds in the same way but is
somewhat simpler as there are fewer divergent terms. The final expression for
the charges is the same, but some of the terms can be dropped as they give
zero. To be precise:
\begin{itemize}
\item $m^2_{\ast} + \frac{81}{100 \,l^2} < m^2< m^2_{\ast}
+\frac{1}{l^2}$ ($1/2<\Delta < 3/5$): all the terms in (\ref{QPhi4}) contribute
to the surface integral and the non linear contribution (\ref{DeltaQ}) is
essential; \item $m^2_{\ast} + \frac{9}{16 \, l^2} < m^2 < m^2_{\ast}
+\frac{81}{100 \, l^2}$ ($3/5<\Delta < 3/4$): the term proportional to $k_{5}$
can be dropped but the non linear contribution (\ref{DeltaQ}) remains
essential; \item $m^2_{\ast} + \frac{1}{4 \, l^2} < m^2 < m^2_{\ast}
+\frac{9}{16 \, l^2}$ ($3/4<\Delta < 1$): both $k_{4}$ and $k_{5}$ can be
dropped, as well as the non linear contribution (\ref{DeltaQ}) to the
gravitational charge; \item $m^2_{\ast}  < m^2 < m^2_{\ast} +\frac{1}{4 \,
l^2}$ ($1<\Delta <3/2$): the terms proportional to $k_{3}$, $k_{4}$, $k_{5}$
and the non linear contribution (\ref{DeltaQ}) are subleading.
\end{itemize}

\subsection{Compatibility with equations of motion}

When imposing boundary conditions at infinity, there is always the danger of
eliminating by hand interesting solutions that would not have the prescribed
fall-off.  We show here that our boundary conditions are compatible with the
equations of motion.  This would not be the case had we not allowed terms that
behave like $r^{- 2 \Delta}$, $r^{-3 \Delta}$ or  $r^{-4 \Delta}$, which are
forced by the non linearities of the field equations. Otherwise, the treatment
would only be valid in the range when these non linear effects are subleading,
i.e., $m^2_{\ast}  < m^2 < m^2_{\ast} +\frac{1}{4 \, l^2}$ ($1<\Delta <3/2$).

We consider the static and spherically symmetric case for simplicity.  The
metric has the form
\begin{equation}
\displaystyle d{s}^{2}=-\left[  1+\frac{r^{2}}{l^{2}}+O(r^{-1})\right]
dt^{2}+\frac{dr^{2}}{1+\frac{r^{2}}{l^{2}}-\frac{\mu(r)}{r}}+r^{2}%
d\Omega^2\;,\label{ds2}%
\end{equation}
where $\mu(r)$ must grow slower than $r^{3}$ for $r\rightarrow\infty$ in order
to preserve the value of the cosmological constant (but it can overcome the 1
in $g_{rr}$). The nonlinear Klein-Gordon equation then reads,
\begin{align}
\left(  \frac{2}{r}+\partial_{r}+\frac{1}{2}\partial_{r}(\log[-g_{tt}%
g_{rr}])\right)  (g^{rr}\partial_{r}\phi)-m^{2}\phi &  =\frac{dU}{d\phi
}\nonumber\\
=l^{-2}\left(  3C_{3}\phi^{2}+4C_{4}\phi^{3}+5C_{5}\phi^{4}+\cdot\cdot
\cdot\right) .  & \label{nonlinKG}%
\end{align}
Expanding $\phi$ as a power series, the leading term of the scalar field is of
the form
\[
\phi(r)=\frac{a}{r^{\Delta}}+\cdots\ ,
\]
as in the linear case. Nonlinearities are felt at the next order and do indeed
arise from  the self-interacting potential if the scalar field mass is large
enough. In this case, the cubic self-interacting term forces the next leading
order to be of the form
\[
\phi(r)=\frac{a}{r^{\Delta}}+\beta_{1}\frac{a^{2}}{r^{2\Delta}}%
\cdots\ ,
\]
in order to match both sides of Eq. (\ref{nonlinKG}).  Here $\beta_1$ is
precisely given by (\ref{Betas-Gravt}) and (\ref{Betas-AdS} (this is in fact
how it might be fixed).

Depending on the scalar field mass, the next relevant orders in the
Klein-Gordon equation can also depend on the next terms of self-interacting
potential {\em as well as on the gravitational back-reaction} through $\mu(r)$
in Eq. (\ref{ds2}), which can be found as a series solving the constraint
$H_{\perp}=0$,
\begin{equation}
\mu^{\prime}+\frac{r}{2}(\phi^{\prime})^{2}\mu=\kappa \frac{r^{2}}{2}\left[
\left( \frac{r^{2}}{l^{2}}+1\right)  (\phi^{\prime})^{2}+m^{2}\phi^{2}+2U(\phi
)\right]  \ .\label{H perp}%
\end{equation}
Substituting the asymptotic form of the scalar field in (\ref{H perp}) provides
a series expression for the back reaction $\mu(r)$ which can be plugged back
into the Klein-Gordon equation (\ref{nonlinKG}) to determine the next terms in
the series of $\phi$. These equations can be solved consistently as a power
series to yield both Eq. (\ref{AsymptPhi4}) and
\begin{equation}
\mu=\frac{\kappa r^{3}}{l^{2}}(\alpha_{1}a^{2}r^{-2\Delta}+\alpha_{2}a^{3}%
r^{-3\Delta}+(\alpha_{3}-\alpha_1 )a^{4}r^{-4\Delta}+(\alpha_{4}-2\alpha_1
\alpha_2)a^{5}r^{-5\Delta})+\mu
_{0}\ ,\label{Mu4}%
\end{equation}
where the $\beta_{i}$  and the $\alpha_{j}$ are the constants given by
(\ref{Betas-Gravt}), (\ref{Betas-AdS}) and (\ref{alphas-charges-4}). This
behavior corresponds precisely to the asymptotic conditions (\ref{AsymptPhi4})
and (\ref{AsymptMetric4}) -- which were derived in fact in this manner. It is
notable that the values of the constants $\beta_i$ and $\alpha_j$ that follow
upon integration of the equations of motion also cancel all divergences in the
surface integrals. In other words, the same results are found solving the
hamiltonian constraints and imposing finiteness of the charges.

It is worth pointing out that for the asymptotic form of the scalar field
(\ref{AsymptPhi4}), the gravitational back reaction merely amounts to a
redefinition of the coefficients $\beta_{2}$ and $\beta_{3}$, and hence, its
effect mimics the nonlinearity of the self-interaction (which are present even
in a pure anti-de Sitter background).  Consequently, even in the absence of a
self interacting potential, the $\beta_2$ term is switched on due to the
gravitational back reaction for a large enough mass. In fact, both effects can
even cancel each other out. For example, the effects of the self interaction
can be completely screened by choosing a particular family of potentials. The
standard asymptotic behavior of the free scalar field in Eq.
(\ref{Asympt-Phi-Linear}) is then recovered by imposing
$\beta_{1}=\beta_{2}=\beta_{3}=0$.  This implies the
following restrictions on the self interaction%
\begin{equation}
U(\phi)=\left\{
\begin{array}
[c]{ll}%
O(\phi^{3}) & :1<\Delta <3/2\\
O(\phi^{4}) & :3/4<\Delta \leq 1\\
-\frac{\kappa\Delta^{2}(3-2\Delta)}{8l^{2}}\phi^{4}+O(\phi^{5}) & :3/5<\Delta
\leq 3/4\\
-\frac{\kappa\Delta^{2}(3-2\Delta)}{8l^{2}}\phi^{4}+O(\phi^{6})\; & :1/2<\Delta
\leq 3/5
\end{array}
\right.\, , \label{U}%
\end{equation}
where the equalities hold even in the presence of logarithmic branches (see
section \ref{logs}).

\section{Higher dimensions}
\setcounter{equation}{0} \label{HigherD}

The analysis becomes simpler in higher dimensions as most of the difficulties
encountered in four dimensions go away.  We only give the results as the
derivation proceeds along identical lines. There are only three cases to be
considered:
\begin{itemize}
\item $D=5, 6$, $m^2_{\ast} + \frac{(D-1)^2}{36 \, l^2} < m^2 <m^2_{\ast} + \frac{1}{l^2}$,
\item $D = 5,6$, $m^2_{\ast} < m^2 < m^2_{\ast} + \frac{(D-1)^2}{36 \, l^2}$,
\item $D \geq 7$, $m^2_{\ast} < m^2 <m^2_{\ast} + \frac{1}{l^2}$,
\end{itemize}
the last two cases being treated similarly.  In the first case, $r^{-\Delta}$
and $r^{-2\Delta}$ (but not $r^{-3\Delta}$) dominate asymptotically over
$r^{-\Delta_+}$.  In the last two cases, only $r^{-\Delta}$ (but not
$r^{-2\Delta}$) dominates asymptotically over $r^{-\Delta_+}$.

\subsection{$D=5, 6$, $m^2_{\ast} + \frac{(D-1)^2}{36 \, l^2} < m^2 <m^2_{\ast} +
\frac{1}{l^2}$}

In this case, the asymptotic conditions for the scalar field and the metric are
given by
\begin{align}
\phi &  =ar^{-\Delta}+\beta_{1}a^{2}r^{-2\Delta}+br^{-\Delta_{+}}+\cdot
\cdot\cdot\ ,\label{AsymptPhiD}\\
& \nonumber
\end{align}
and
\begin{equation}%
\begin{array}
[c]{lll}%
h_{rr} & = & \displaystyle\frac{\kappa l^{2}}{r^{2}}\left(  \alpha_{1}a^{2}%
r^{-2\Delta}+\alpha_{2}a^{3}r^{-3\Delta}\right)  +\frac{f_{rr}}{r^{(D+1)}%
}\\[3mm]%
h_{mn} & = & \!\displaystyle\frac{f_{mn}}{r^{(D-3)}}+\cdots\\[3mm]%
h_{mr} & = & \!\displaystyle\frac{f_{mr}}{r^{(D-2)}}+\cdots
\end{array}
\label{AsymptMetricD}%
\end{equation}
where the indices $m,n$ are purely angular and where, as in the four
dimensional case $b$, $f_{rr}$ $f_{mn}$ and $f_{rm}$ are independent functions
of time and the angles ($x^{m}$).  The coefficients $\beta_{1}$, in the scalar
field, and $\alpha_{1}$ and $\alpha_{2}$ in the metric, are constants given by
\begin{equation}
\beta_{1}=\frac{C_{3}}{\Delta(\Delta-(D-1)/3)}\ ,\label{Beta1D}%
\end{equation} and
\begin{equation}
\alpha_{1}    =-\frac{\Delta}{D-2}\, , \; \; \; \alpha_{2}
=-\frac{8\Delta}{3(D-2)}\beta_{1}
\ .\label{alphas-charges-D}%
\end{equation}
The conjugate momenta fulfill
\begin{align}
\pi^{rr}=O(r),\quad &  \pi^{rm}=O(r^{-2}),\quad\pi^{mn}=O(r^{D-2\Delta
-6}),\label{PiGD}\\
&  \pi_{\phi}=O(r^{D-4-\Delta})\,,\label{PiPhiD}%
\end{align}
{}Finally, $a$ is fixed in terms of $b$ as
\begin{equation}
a=a_{0}b^{\frac{\Delta}{\Delta_{+}}}\ ,\label{abDAdS}%
\end{equation}
where $a_{0}$ is an arbitrary dimensionless constant. Just as in the
four-dimensional case, the existence of a relationship between $a$ and $b$ is
necessary in order to get integrable charges.  That relationship is then fixed
to be of the form (\ref{abDAdS}) by anti-de Sitter invariance.

Again, the asymptotic behavior of the metric acquires a strong back reaction in
comparison with the standard fall-off Eq. (\ref{Standard-Asympt}). Unlike the
four-dimensional case, the gravitational back-reaction has, however, no
influence in the asymptotic form of the scalar field.

\subsection{$D = 5,6$, $m^2_{\ast} < m^2 < m^2_{\ast} + \frac{(D-1)^2}{36 \, l^2}$
and $D \geq 7$}

The boundary conditions are then simpler and read
\begin{align}
\phi &  =ar^{-\Delta}+br^{-\Delta_{+}}+\cdot
\cdot\cdot\ ,\label{AsymptPhiD'}\\
& \nonumber
\end{align}
and
\begin{equation}%
\begin{array}
[c]{lll}%
h_{rr} & = & \displaystyle\frac{\kappa l^{2}}{r^{2}}\left(  \alpha_{1}a^{2}%
r^{-2\Delta}\right)  +\frac{f_{rr}}{r^{(D+1)}%
}\\[3mm]%
h_{mn} & = & \!\displaystyle\frac{f_{mn}}{r^{(D-3)}}+\cdots\\[3mm]%
h_{mr} & = & \!\displaystyle\frac{f_{mr}}{r^{(D-2)}}+\cdots
\end{array}
\label{AsymptMetricD'}%
\end{equation}
with
\begin{equation} \alpha_{1}    =-\frac{\Delta}{D-2} \, , \; \; \; \;
a=a_{0}b^{\frac{\Delta_{-}}{\Delta_{+}}}\ .\label{abDAdS'} \end{equation}

In this case, the self-interacting potential has no effect on the asymptotic
form of the scalar field, which coincides with the one obtained for the linear
approximation in a fixed AdS background as in Eq. (\ref{Asympt-Phi-Linear}).

\subsection{Symmetries and Generators}
The asymptotic conditions given above are preserved under the asymptotic AdS
symmetry, and the functional relation between $a$ and $b$ required for the
integrability of the symmetry generators, acquires the same form as in the four
dimensional case given by Eq. (\ref{ab4AdS}).

Following Ref. \cite{Regge-Teitelboim}, one can compute the charges. With our
boundary conditions, the divergences cancel and the charges are found to be
\begin{equation}
Q(\xi)=Q_{0}(\xi)+Q_{\phi}(\xi)\;,\label{Q-Tot-D}%
\end{equation}
where $Q_{0}$ is given by the standard formula in Eq. (\ref{Q0}), and
\begin{equation}
Q_{\phi}=\frac{1}{2(D-1)l}\int d^{2}\Omega r^{(D-2)}\xi^{\perp}\left[
l^2(n^r\partial_r\phi)^{2} -m^{2}l^{2}\phi^{2}+k_{3}\phi^{3}\right]
\ ,\label{QPhiD}%
\end{equation}
with
\begin{equation} \label{k3}
k_{3}=-2C_3 .
\end{equation}
The term proportional to $k_{3}$ is needed only for $D=5,6$ and $m^2_{\ast} +
\frac{(D-1)^2}{36 \, l^2} \leq m^2 < m^2_{\ast} + 1$. (As seen in the previous
section, this term is also necessary in the range $m^2_*+\frac{1}{4l^2}< m^2<
m^2_* + \frac{9}{16l^2}$, and for larger values of $m^2$, the terms
proportional to $k_4$ and $k_5$ are also necessary.) Note that Eq.
(\ref{QPhiD}) can then be extrapolated to the case when the BF bound is
saturated, and it can also be seen to hold for $\Delta_+=2\Delta$. The algebra
of the charges (\ref{Q-Tot-D}) coincides with the standard one, i. e., the AdS
algebra $SO(D-1,2)$.

It is also worth noting that in dimensions higher than four, the gravitational
back reaction cannot mimic the nonlinearity of the self interaction, so that
the ``screening" effect observed in section 4.7 above is absent.

\section{Logarithmic terms of nonlinear origin}
\setcounter{equation}{0} \label{logs}

In general, for any dimension, logarithmic branches are present when
$\frac{\Delta_+}{\Delta}=n$ is a positive integer. In this case the scalar
field acquires a logarithmic branch of the form
$$
\phi = ar^{-\Delta} + \cdots +h a^n r^{-\Delta_+} \log(r) + br^{-\Delta_+} +
\cdots,
$$
where $h$ is a fixed constant explicitly determined below.

The critical values of the spacetime dimensions and mass for which this
phenomenon occurs are
\begin{itemize}
\item $D=4, 5$, $6$, $m^2= m^2_{\ast} + \frac{(D-1)^2}{36 l^2}, $ ($n=2$),
\item $D = 4$, $m^2 = m^2_{\ast} + \frac{9}{16 \, l^2},$ ($n=3$),
\item $D =4$, $m^2 = m^2_{\ast} + \frac{81}{100 l^2},$ ($n=4$).
\end{itemize}

As in the generic case, integrability of the charges requires $a$ and $b$ to be
functionally related. The asymptotic AdS symmetry implies a functional relation
given by
\begin{equation}
b=a^n\left[ b_0 -\frac{h}{\Delta} \log a\right].
\end{equation}
which is different from the generic form (\ref{abDAdS}) due to the presence of
the logarithmic branch. Note that this relation also holds for $n=1$, when the
scalar field saturates the BF bound.

\subsection{$D=4,5,6$, $m=m^2_*+ (D-1)^2/(36l^2)$}

In this case, $\Delta_+=2\Delta=\frac{2(D-1)}{3}$, and the asymptotic behavior
of the scalar field and the metric is given by

\begin{equation}%
\begin{array}
[c]{lll}%
\phi & = & \displaystyle \frac{a}{r^{\Delta}} -\frac{9 C_3}{D-1} a^2\frac{\log
r}{r^{2\Delta}} + \frac{b}{r^{2\Delta}} + ... \\
h_{rr} & = & \displaystyle\frac{\kappa l^{2}}{r^{2}}\left(  \alpha_{1}a^{2}%
r^{-2\Delta} + \frac{8C_3}{D-2}a^3 r^{-3\Delta} \log r \right)  +\frac{f_{rr}}{r^{D+1}%
},\\[3mm]%
h_{mn} & = & \!\displaystyle\frac{f_{mn}}{r^{D-3}}+\cdots,\\[3mm]%
h_{mr} & = & \!\displaystyle\frac{f_{mr}}{r^{D-2}}+\cdots,
\end{array}
\label{Log456}%
\end{equation}
where $\alpha_1=-\Delta/2$, as in the generic case. Proceeding as in the
generic case for these asymptotic conditions, it is found that the divergences
cancel out, and the charges are still expressed in the form (\ref{Q-Tot-D}) and
(\ref{QPhiD}). The relationship between $a$ and $b$ required by asymptotic AdS
symmetry is now given by
\begin{equation}\label{log3}
b=a^2\left[ b_0 + \frac{27C_3}{(D-1)^2} \log a\right].
\end{equation}

Note that as the mass approaches the critical value $m^2_*+ (D-1)^2/(36l^2)$
from above, the coefficient $\beta_1$ in Eq. (\ref{Beta1D}) develops a pole at
$\Delta=(D-1)/3$. Thus in the limit, the following replacement takes place:
\begin{equation}
\frac{\beta_1}{r^{2\Delta}} \rightarrow (D-1-3\Delta)\beta_1\frac{\log
r}{r^{2\Delta}}.
\end{equation}
Also to be pointed out is the fact that the logarithmic terms are absent if
$C_3 = 0$.

\subsection{$D = 4$, $m^2 = m^2_{\ast} + \frac{9}{16\, l^2}$}

In this case, $\Delta_+=3\Delta =9/4$. The asymptotic behavior of the scalar
field and the metric is
\begin{equation}
\begin{array}
[c]{lll}%
\phi & = & \displaystyle \frac{a}{r^{\Delta}} +\frac{\beta_1a^2}{r^{2\Delta}}+
(3-4\Delta)\beta_2a^3\frac{\log r}{r^{3\Delta}} + \frac{b}{r^{3\Delta}} + ...
\\
h_{rr} & = & \displaystyle\frac{\kappa l^{2}}{r^{2}}\left(  \alpha_{1}a^{2}%
r^{-2\Delta} + \alpha_2 a^3 r^{-3\Delta} + (4\Delta-3)\beta_2a^4
\frac{9\log r}{8r^{4\Delta}} \right)  +\frac{f_{rr}}{r^5}\\[3mm]%
h_{mn} & = & \!\displaystyle\frac{f_{mn}}{r}+\cdots\\[3mm]%
h_{mr} & = & \!\displaystyle\frac{f_{mr}}{r^2}+\cdots
\end{array}
\label{Log4456}%
\end{equation}
where $\beta_1$ and $\beta_2$ are given by (\ref{Betas-Gravt}), and $\alpha_1$
and $\alpha_2$, are those of (\ref{alphas-charges-4}), as in the generic case.
The divergences of the charges cancel, and they are still expressed as $Q= Q_G
+ Q_{\phi}$, where the term proportional to $k_5$ in (\ref{QPhi4}) is
subleading, but the nonlinear contribution to the gravitational part of the
charge $\Delta Q$ given by (\ref{DeltaQ}) is relevant.

The relationship between $a$ and $b$, required by asymptotic AdS symmetry, is
now given by
\begin{equation}\label{log4}
b=a^3\left[ b_0 + \frac{(4\Delta-3)\beta_2}{\Delta} \log a\right],
\end{equation}
where the factor $(4\Delta-3)$ cancels the pole of $\beta_2$ at $\Delta=3/4$.

\subsection{$D=4$, $m^2=m^2_*+\frac{81}{100 l^2}$}

Now $\Delta_+=4\Delta =12/5$, and the asymptotic behavior of the fields is
given by
\begin{equation}
\begin{array}
[c]{lll}%
\phi & = &  \displaystyle \frac{a}{r^{\Delta}} +\frac{\beta_1a^2}{r^{2\Delta}}+
\frac{\beta_2a^3}{r^{3\Delta}} - (5\Delta -3)\beta_3 a^4\frac{\log
r}{r^{4\Delta}} + \frac{b}{r^{4\Delta}} + ...
\\
h_{rr} & = & \frac{l^2}{r^2}\left[ \alpha_1a^2 r^{-2\Delta} +\alpha_2 a^3
r^{-3\Delta} + \alpha_3 a^4 r^{-4\Delta}+ \frac{24}{25 r^{5\Delta}}(5\Delta
-3)\beta_3a^5\log r\right] +f_{rr}/r^5.\\[3mm]%
h_{mn} & = & \!\displaystyle\frac{f_{mn}}{r}+\cdots\\[3mm]%
h_{mr} & = & \!\displaystyle\frac{f_{mr}}{r^2}+\cdots
\end{array}
\label{Log44456}%
\end{equation}
where $\beta_1$, $\beta_2$ and $\beta_3$ are given by (\ref{Betas-Gravt}), and
$\alpha_1$, $\alpha_2$ and $\alpha_3$, are those of (\ref{alphas-charges-4}),
as in the generic case. The divergences of the charges again cancel, and they
are given by $Q= Q_G + Q_{\phi}$, where the all the terms in (\ref{QPhi4}), as
well as the nonlinear contribution to the gravitational part of the charge
$\Delta Q$ given by (\ref{DeltaQ}) are relevant.

The relationship between $a$ and $b$, required by asymptotic AdS symmetry, is
now given by
\begin{equation}\label{log5}
b=a^4\left[ b_0 + \frac{(5\Delta-3)\beta_3}{\Delta} \log a\right],
\end{equation}
where the factor $(5\Delta-3)$ cancels the pole of $\beta_3$ at $\Delta=3/5$.

\section{Remarks on non-AdS invariant boundary conditions}
\setcounter{equation}{0} \label{breaking}

\subsection{Breaking AdS invariance through the boundary conditions on the scalar
field}

The existence (integrability) of the charges (in particular, the energy) forces
$a$ and $b$ to be functionally related in a way that is essentially unique if
one insists on AdS invariance. However, one may consider different functional
relationships.  In that case, although the metric still has the same asymptotic
AdS invariance, the scalar field breaks the symmetry to $\mathbb{R}\times
SO(D-1)$ because the asymptotic form of $\phi$ is not maintained under the
action of $\xi^{r}$. This breaking of asymptotic AdS invariance has been
considered in \cite{solitons,HHollands}, following ideas from the AdS/CFT
correspondence \cite{Witten}. One may still develop the formalism provided
that, as above, one takes proper account of the non linearities in the
equations when these are relevant.

It is worth pointing out that requiring the matter piece of the charges
$Q_{\phi}$ to be integrated as an analytic local functional of $\phi$ and its
derivatives,
\begin{equation}
Q_{\phi}=\int_{S^{D-2}}\sqrt{h}d^{D-2}x\xi^{\perp}F(\phi,n^{i}\partial_{i}%
\phi,n^{i}n^{j}\partial_{i}\partial_{j}\phi,...),\label{Q-phi}%
\end{equation}
where $F$ is a polynomial in its entries and $n^{i}$ is a unit normal to the
sphere at infinity, is strong enough to fix the relation between $a$ and $b$
within a one-parameter family. In the generic case ($\Delta_+/\Delta\neq n$),
this relation is of the form
\begin{equation}
\displaystyle a=a_{0}b^{\frac{k_{1}\Delta}{(D-1)-k_{1}\Delta}}%
\; ,\label{abLocalityD}%
\end{equation}
where $k_1$ is some constant (equal to 1 in the asymptotically anti-de Sitter
invariant case). The case $\Delta_+/\Delta= n$ will be discussed at the end of
this subsection.

This can be seen as follows. Since the field satisfies a second order equation,
it is expected that $F$ should depend only on $\phi$ and $\partial_{r}\phi$ at
infinity. In fact, one can see that, using the asymptotic conditions
(\ref{AsymptPhiD}), the higher derivatives terms can always be expressed as
linear combinations of $\phi$ and $n^r\partial_r\phi$. The leading terms are
then $\phi^{2}$, $(n^r\partial_r\phi)^2$ and $\phi n^r\partial_r\phi$, and by
virtue of the asymptotic conditions, without loss of generality one can drop
the term\footnote{There is always a precise linear combination of these three
terms that give no contribution to the charge $Q_{\phi}$, except for $m^2
=m^2_*$.} $\phi n^r\partial_r\phi$. In consequence, for $D\geq5$ dimensions,
requiring the variation of this local functional to match what one gets from
the Hamiltonian constraint fixes the form of the charge to be
\begin{equation}
Q_{\phi}^{k_{1}}=\frac{1}{2(D-1)l} \int d^{D-2}\Omega r^{(D-2)}\xi^{\perp}
\left[  k_{1}l^2(n^r\partial_r\phi)^{2}+k_{2}\phi^{2}+k_{3}\phi^{3}\right]  \ ,
\end{equation}
with
\begin{align}
k_{2}  &  =\Delta(D-1-\Delta k_{1})\ ,\\
k_{3}  &  =2\beta_{1}\Delta\left(  \frac{D-1}{3}-k_{1}\Delta\right),
\end{align}
and requires $a $ and $b$ to be related as in Eq. (\ref{abLocalityD}). It is
easy to see that for $D\geq7$, the cubic term is subleading. Note that for
$k_1=1$, the expressions for the asymptotically AdS invariant case are
recovered c.f., (\ref{QPhiD}), (\ref{k3}).

In four dimensions, higher order terms are needed, so that the matter piece of
the charge reads
\begin{equation} \label{Qk1}
Q_{\phi}^{k_{1}}=\frac{1}{6l}\int d^{2}\Omega r^{2}\xi^{\perp}\left[
k_{1}l^2(n^r\partial_r\phi)^2+k_{2}\phi^{2}+k_{3}\phi^{3}+k_{4}\phi^{4}%
+k_{5}\phi^{5}\right]  \ ,
\end{equation}
where
\begin{eqnarray}
k_{2}  & =& (3-k_{1}\Delta)\Delta\\
k_{3}  & =& 2\beta_{1}\Delta(1-k_{1}\Delta)\\
k_4    & = & \Delta_-[3\beta_1^2(k_1 \Delta_--1) -
(\beta_2 + \kappa\frac{\Delta_-}{8})(4k_1 \Delta_--3)]\\
k_5    & = & \Delta_-[\frac{6 \beta_3}{5}(3-5k_1 \Delta_-)- 6 \beta_1^3(k_1
\Delta_- -1)  \nonumber \\
& & + \kappa \frac{\beta_1 \Delta_-}{6}(3-8k_1 \Delta_-) +\frac{12\beta_1
\beta_2}{5}(5k_1 \Delta_- -4)].
\end{eqnarray}
For $k_1=1$ the expressions for the asymptotically AdS invariant case,
(\ref{QPhi4}) and (\ref{ks}), are recovered.

For the cases in which the fields develop logarithmic branches,
$\Delta_+/\Delta=n$, the results can be summarized as follows.

\begin{itemize}
\item $D=4, 5$, $6$, $m^2= m^2_{\ast} + \frac{(D-1)^2}{36 l^2}$. The charge
is given by (\ref{Qk1}), where $k_1=1$ and $k_2=-m^2 l^2$, as for
asymptotically AdS case, but now instead $k_3$ is arbitrary, labelling the
relationship between $a$ and $b$,
\begin{equation}\label{log3-}
b=a^2\left[ b_0 - \frac{27k_3}{2(D-1)^2} \log a\right].
\end{equation}
For $k_3=-2C_3$ the expression (\ref{log3}) is recovered, so that the
asymptotic AdS symmetry is restored, and all $k$'s are as in the generic case
for AdS (\ref{QPhiD}).
\item $D = 4$, $m^2 = m^2_{\ast} + \frac{9}{16 \, l^2}$. The charge is given
by (\ref{Qk1}), with $k_1=1$, $k_2=-m^2 l^2$, and $k_3=-2C_3$ (as for AdS), but
now $k_4$ is arbitrary, and the $k_5$ term is irrelevant. In this case, the
relationship between $a$ and $b$ is given by
\begin{equation}\label{log4-}
b=a^3\left[ b_0 - \frac{16 k_4+9\beta_1^2}{9} \log a\right].
\end{equation}
The asymptotic AdS symmetry is recovered for $k_4 =  -2C_4 - \kappa
\frac{27}{128}$, in agreement with (\ref{log4}) and (\ref{ks}).

\item $D =4$, $m^2 = m^2_{\ast} + \frac{81}{100 l^2}$. The charge is given
by (\ref{Qk1}), where all the terms are relevant. In the general case (when AdS
symmetry is broken by the scalar field), $k_1$, $k_2$, $k_3$, and $k_4$ are
fixed as in (\ref{ks}) (as for AdS), but now $k_5$ is arbitrary. In this case,
the relationship between $a$ and $b$ is given by
\begin{equation}\label{log}
b=a^4\left[ b_0 +(
-\frac{125k_5}{54}-\kappa\frac{\beta_1}{4}+\frac{10\beta_1^3}{3}-
\frac{10\beta_1\beta2}{3})\log a\right].
\end{equation}

The asymptotic AdS symmetry is recovered for $k_5 = -2C_5 +\kappa \frac{3
C_3}{4} $, in agreement with (\ref{ks}), and (\ref{log5}).

\item $D\geq3$, $m^2=m_*^2$. When the BF bound is saturated, $Q_{\phi}^{k_{1}}$
has the form
\begin{equation}
Q_{\phi}^{k_{1}}=\frac{1}{2(D-1)l} \int d^{D-2}\Omega r^{(D-2)}\xi^{\perp}
\left[k_{0}l^2\phi n^r\partial_r\phi+
k_{1}l^2(n^r\partial_r\phi)^{2}+k_{2}\phi^{2} \right] \ ,
\end{equation}
with $k_0= (k_1-1)(D-1)$ and $k_2=-m_*^2 l^2 k_1$. The relation between $a$ and
$b$ is
$$
b= -k_1 \frac{2}{D-1} a \log a/a_0.
$$
This means that for $k_1=1$ our previous results (\ref{ablog}) and
(\ref{Qphilog}), which are compatible with AdS symmetry, are recovered.
\end{itemize}
For $k_1\neq1$, the total charges $Q=Q_{G}+Q_{\phi}^{k_{1}}$ are finite and
generate the asymptotic symmetry group $\mathbb{R}\times SO(D-1)$. \\

\subsection{Locally asymptotically anti-de Sitter space}

The surface integrals expressing the conserved charges presented here can be
readily extended to configurations where the asymptotic AdS symmetry is broken
through non trivial topology. For instance, the exact four-dimensional black
hole solution of Ref. \cite{MTZTop} which is dressed with a minimally coupled
scalar field with a slow fall-off, has broken rotational invariance in the
asymptotic region since the boundary of the spacelike surface does not
correspond to a sphere in that region \footnote{It was shown in \cite{QN} that
the perturbative stability of locally AdS spacetimes with this kind of topology
holds provided the mass satisfies the same BF bound.}. Thus, in order to
compute the conserved charges for the remaining asymptotic symmetries for this
kind of objects, it is enough to replace the volume element $d\Omega^{D-2}$of
the sphere $S^{D-2}$, by the volume element $d\Sigma^{D-2}$ of the boundary of
the spacelike surface. It is simple to check that the mass for the black hole
in Ref. \cite{MTZTop} can be reproduced in this way.

\section{Conclusions}
\setcounter{equation}{0} \label{conclusions}

In this paper, we have investigated the asymptotics of anti-de Sitter gravity
minimally coupled to a scalar field with a slow fall-off ($a \not= 0$). The
scalar  field  gives rise to a back reaction that modifies the asymptotic form
of the geometry, which is consistent with asymptotically AdS symmetry for
suitable boundary conditions. In turn, additional contributions to the charges,
which are not present in the gravitational part and which depend explicitly on
the matter fields at infinity, arise and insure finiteness of the charges.

The discussion has been carried out here for a minimally coupled
self-interacting scalar field in dimensions $D\geq4$ with any mass between the
Breitenlohner-Freedman bound and the Breitenlohner-Freedman bound plus $1/l^2$.

We have shown that one can consistently include the slow fall off of the scalar
in the Hamiltonian formulation provided a functional relationship is imposed on
$a$ and $b$ ensuring integrability of the charges. We considered only one
scalar field. In the presence of many scalar fields, one must make $a_i \delta
b_i$ integrable, forcing $a_i = \delta L / \delta b_i$ for some $L$. This is in
line with the AdS-CFT correspondence where such functional relationships
(defining ``Lagrangian submanifolds") have been considered in the context of
multi-trace deformations \cite{Witten,Aharony}.

We have also observed that when a non-trivial potential is considered, the
asymptotic form of the scalar field obtained through the linear approximation
is no longer reliable and acquires extra contributions when the mass of the
scalar field is within the range $m_{\ast}^{2}+(D-1)^2/(36 l^2) < m^2 <
m_{\ast}^2 + 1/l^2$ for $D<7$ dimensions.   The effects of the self-interaction
are absent only for a particular class of potentials. The four-dimensional case
is particularly interesting since gravitational back reaction is so strong that
it can even mimic the nonlinearity of the self interaction.  Both effects are
present but can cancel each other out for fine-tuned potentials within a
particular class. Furthermore, the purely gravitational contribution to the
charges acquires an extra term which is nonlinear in the deviation from the
background. These effects were first observed in the three-dimensional case for
an specific value of the scalar field mass \cite{HMTZ1}.

A somewhat unexpected outcome of our analysis is that at the critical values of
the mass where new terms in the potentials become relevant, the
self-interactions of the scalar field as well as its gravitational back
reaction, (not discussed in previous treatments), force the fields to develop
extra logarithmic branches.

One of the advantages of allowing the $a$-branch of the scalar field -- and
hence, considering the relaxed asymptotic behavior discussed here -- is that
the space of physically admissible solutions is then enlarged and includes new
interesting objects. In particular, asymptotically AdS black hole solutions
having scalar hair with slow fall-off have been found numerically in Refs.
\cite{hairy,Hertog-Maeda1, Hertog-Maeda2,Numerical-BHs}. Numerical black hole
solutions exhibiting these features have also been found for non-abelian gauge
fields with a dilatonic coupling in Ref. \cite{Radu-Tchrakian}.

The possibility to consider the two scalar branches simultaneously is a feature
peculiar to anti-de Sitter space, which is not available when the cosmological
constant vanishes.  In that case, one can only include the decaying mode $\sim
exp(- m r)/r$ at infinity.

Another interesting limit is obtained when one switches off the gravitational
coupling constant. In this case, the matter piece of the charge reduces to
$$
Q^{AdS}_{\phi}= \frac{l}{D-1}\int_{\partial \Sigma} d^{D-2}\Omega
r^{(D-2)}\xi^{\perp}[(n^r\partial_r \phi)^{2}/2-m^2\phi^2/2-U(\phi)].
$$
This boundary term is sufficient to regularize the generators on a fixed AdS
background as
$$
G^{AdS}(\xi)= \int_{\Sigma} \xi^{\mu} T_{\mu \nu} dS^{\nu} + Q^{AdS}_{\phi}.
$$
In this way, there is no need to invoke an "improvement" coming from the non
minimal coupling between gravity and the scalar field.

The effect of the relaxed asymptotic behavior discussed here opens new
questions that deserve further study, as for instance, the positivity of the
energy in this wider context, its compatibility with supersymmetry, as well as
its holographic significance\footnote{While this paper was finished, we have
been informed  by Max Ba\~nados, Adam Schwimmer and Stefan Theissen about a
holographic interpretation of the logarithmic relations between the boundary
conditions found in \cite{HMTZ2}.}. Incidentally, the critical values of the
mass where a new term in the potential, say $\phi^k$, becomes relevant,
corresponds precisely to the case where the $k$-th power of the dual field
(which has dimension $\Delta$) becomes relevant in the sense of
the dual conformal field theory.  A related question is to derive
the above charges through holographic methods \cite{Sken}.\\

\textbf{Acknowledgments} We thank Aaron Amsel, Riccardo Argurio, Max Ba\~nados,
Eloy Ay\'{o}n-Beato, Claudio Bunster, Gast\'{o}n Giribet, Stanislav Kuperstein,
Don Marolf, Herman Nicolai, Rub\'{e}n Portugu\'{e}s, Adam Schwimmer and Stefan
Theisen,  for useful discussions and enlightening comments. We are also
grateful to Geoffrey Comp\`ere for spotting an important typo in the first
version. This work was funded by an institutional grant to CECS of the
Millennium Science Initiative, Chile, and Fundaci\'on Andes, and also benefits
from the generous support to CECS by Empresas CMPC. The work MH is partially
supported by IISN - Belgium (convention 4.4505.86), by the ``Interuniversity
Attraction Poles Programme -- Belgian Science Policy " and by the European
Commission programme MRTN-CT-2004-005104, in which he is associated to V.U.
Brussel. This research was also supported in part by FONDECYT grants N${{}^o}$
1020629, 1040921, 1051056, 1051064, 1061291, 7050232 as well as by the National
Science Foundation under Grant No. PHY99-07949.

\end{document}